\documentclass[preprint]{elsarticle}

\usepackage[dvipsnames]{xcolor}

\usepackage{booktabs} % For formal tables
\usepackage{listings}
\usepackage{caption}
\usepackage{graphicx}
\usepackage{subcaption}
\usepackage{amsmath}
\usepackage{amssymb}
\usepackage{framed}
\usepackage{multirow}

\usepackage{xcolor}
\usepackage{pgfplots}
\usepackage{pgfplotstable}
\pgfplotsset{compat=1.11,
        /pgfplots/ybar legend/.style={
        /pgfplots/legend image code/.code={%
        \draw[##1,/tikz/.cd,bar width=3pt,yshift=-0.2em,bar shift=0pt]
plot coordinates {(0cm,0.8em)};},
},
}

\usepackage{tikz}

\definecolor{dgreen}{HTML}{219601}
\definecolor{bblue}{HTML}{4F81BD}
\definecolor{rred}{HTML}{C0504D}
\definecolor{oorange}{HTML}{ffa435}
\definecolor{ggreen}{HTML}{9BBB59}
\definecolor{dred}{HTML}{5c0802}

\usepackage{multirow}
\usepackage{mwe}

\usepackage{epsfig}
\usepackage{amsmath}
\usepackage{color}

\usepackage{hyperref}

\graphicspath{ {img/} }

\begin{document}
%%%%%%%%%%%%%%%%%%%%%%%%%%%%%%%%%%%%%%%
%%%%%%%%% TITLE %%%%%%%%%%%%%%%%%%%%%%%
%%%%%%%%%%%%%%%%%%%%%%%%%%%%%%%%%%%%%%%

\title{Personalised Novel and Explainable \\ Matrix Factorization}

\author{Ludovik Coba, Panagiotis Symeonidis, Markus Zanker}

\address{Free University of Bolzano, Italy \\ lucoba@unibz.it, psymeonidis@unibz.it, mzanker@unibz.it}
%\address[mysecondaryaddress]{psymeonidis@unibz.it}

\begin{abstract}
Recommendation systems personalise suggestions to individuals to help them in their decision making and exploration tasks. In the ideal case, these recommendations, besides of being accurate, should also be novel and explainable. However, up to now most platforms fail to provide both, novel recommendations that advance users' exploration along with explanations to make their reasoning more transparent to them. For instance, a well-known recommendation algorithm, such as matrix factorisation (MF), optimises only the accuracy criterion, while disregarding other quality criteria such as the explainability or the novelty, of recommended items. In this paper, to the best of our knowledge, we propose a new model, denoted as NEMF,  that allows to trade-off the MF performance with respect to the criteria of  novelty and explainability, while only minimally compromising on accuracy. In addition, we recommend a new explainability metric based on nDCG, which distinguishes a more explainable item from a less explainable item. An initial user study indicates how users perceive the different attributes of these ``user'' style explanations and our extensive experimental results demonstrate that we attain high accuracy by recommending also novel and explainable items. 
\end{abstract}

\maketitle

% blue - Ludovik

\section{Introduction}

Recommender systems aim primarily at providing accurate item recommendations while ignoring many times additional quality criteria such as the novelty of a recommended item~\cite{Castells2015} or the system's ability to be able to explain to users why specific items are recommended~\cite{Jannach:2016:BeyondMatrix}.

As far as explainability is concerned, Abdollahi and Nasraoui~\cite{Abdollahi:2016:EMF:2872518.2889405, Abdollahi:2017} recently proposed Explainable Matrix Factorization (EMF), where recommendations are not only optimised according to their presumed accuracy but also based on their explainability to users. However, they have not considered the novelty aspect, which risks to recommend explainable items because they are just popular. Such recommendations can be expected items, which might already be known to the user and finally resulting in trivial recommendations. 
%Explainability focuses on user-style explanations, where more ratings for the proposed items among the user's most similar neighbors means higher explainability. 
 
As far as novelty is concerned,  there are many definitions~\cite{Castells2015}. For instance, popularity-based novelty focuses on discovering non-popular products that match the crowd's interest. However, this definition of novelty does not capture how novel an item is for an individual. In the context of traditional MF to provide novel item recommendations, related work~ \cite{Cremonesi2010, Yin_VLDB2012}  observed that by raising the dimensionality of the MF model (i.e., by increasing the number of latent factors), we can more accurately recommend items coming from  the long tail (i.e. more novel items). However, please note that this can seriously harm the efficiency of the MF model.

In this paper, inspired by the work of both ~\cite{Cremonesi2010, Yin_VLDB2012}, who measured the novelty of item recommendations using MF, and Abdollahi and Nasraoui~\cite{Abdollahi:2017}, who proposed Explainable MF, we provide simultaneously personalised novel and explainable item recommendations based on matrix factorization. To the best of our knowledge, we are the first to propose an MF method that accurately recommends simultaneously novel and explainable items.

In terms of explainability, we improve the EMF~\cite{Abdollahi:2017} by reformulating its formula, which now overcomes the problem of Euclidean distance's metric over high dimensional spaces, since it does not put more emphasis on outliers, which may dominate other smaller weights computed for the other data points. Furthermore, we have noticed that the Mean Explainability Precision (MEP)~\cite{Abdollahi:2017}  metric  is not adequate for measuring explainability, because it does not takes the exact rank of the recommended and explainable items into consideration. Instead, we build a new metric, denoted as explainable-nDCG, which is based on the well-known nDCG, and distinguishes a more explainable item from a less explainable item.  

In terms of novelty of a recommended item, since the concept of explainability is based on many ratings among peers, and, thus, it correlates with item popularity~\cite{Cremonesi2010, Yin_VLDB2012}, we propose an algorithmic framework to trade-off between explainability and novelty in matrix factorisation. Our proposed method, denoted as personalised NEMF, has the advantage of controlling through a regularisation term how novel items will be recommended, without having to increase the number of latent factors of MF, which can harm the efficiency of the MF model. Furthermore please note, that item novelty should not be confused with the diversity of recommendation lists, which is later in the discussion section.

In the remainder of this paper, Section~\ref{sec:related} discusses the related work. In Section~\ref{sec:measuring}, we define a new metric for explainability based on the well-known nDCG. Section~\ref{sec:userstudy} describes our findings from a user study so that we consider them in our proposed method. Next, in Section~\ref{objfun} we  propose a framework for personalized novel and explainable MF. Section ~\ref{sec:results} presents and discusses our experimental results on two well-known datasets.  Finally, Section~\ref{sec:conclusions} concludes the paper and describes future work.

\section{Related Work}
\label{sec:related}

The utility of a recommender systems cannot be measured by solely considering the accuracy of recommendations. Jannach et al.~\cite{Jannach:2016:BeyondMatrix},  mentions that additional system aspects, which heavily impact the user experience like explanations, novelty and serendipity gain more attention.

\subsection{Explainability}

Several works discussed explanations for recommender systems up to date. Friedrich and Zanker~\cite{friedrich2011m}, for instance, proposed a taxonomy to classify different approaches to generate explanations for recommendations. According to their taxonomy the explanations we consider in this work are categorised as collaborative explanations, i.e., explanations that justify recommendations based on the amount as well as the concrete values of ratings that derive from similar users, where similarity is typically determined based on similar behaviour and preference expressions during past interactions. The explanation taxonomy proposed by ~\cite{papadimitriou2012generalized} extends this classification by making a distinction based on the three fundamental concepts used for explaining recommendations which are \emph{users}, \emph{items} and item \emph{features}. They can be used to denote the following explanation styles: (i) \emph{User} Style, which provides explanations based on similar users, (ii) \emph{Item} style, which is based on choices made by users on similar items, and (iii) \emph{Feature} Style, which explains the recommendation based on item features (content). Please note, that any combination of the aforementioned styles is then categorised as a multi-dimensional hybrid explanation style. For the ``User" Style, several collaborative filtering recommender systems,
such as Amazon, adopted the following style of justification:
``Customers who bought item $X$ also bought items $Y, Z, \ldots$". This is the so called ``User" style~\cite{BM05} of justification, which is based on users performing similar actions like buying or rating items. When the ``Item"
style of explanation is concerned, justifications are of the form: ``Item $Y$ is
recommended because you highly rated or bought item
$X, Z, \ldots$". Thus, the system depicts those items i.e., $X, Z, \ldots$, that influenced the recommendation of item $Y$ the most. Bilgic et al.~\cite{BM05} claimed that the Item style is preferable over the User style, because it allows users to accurately formulate their true opinion of an item. In case of ``Feature style'' explanations the description of items are exploited to determine a match between a current recommended item and observed user's interests. For instance, restaurants may be described by features such as location, cuisine and cost. Now, if a user has demonstrated a preference for Chinese cuisine and Chinese restaurants are recommended, then explanations will note the Chinese cuisine or their cost aspect. As part of this work we tested users' preference for different explanation styles in a pre-study and determined that the so-called \emph{User} Style (or collaborative user-based ~\cite{friedrich2011m}) explanations scored highest and therefore focused only on this category of explanations for our algorithm development. However, we also discuss if and how our proposed method and metrics can be applied to build or test the quality of other explanation styles in the Discussion section. 

\color{black}

\subsection{Novelty}

\vspace{10pt}

As far as the item novelty is concerned, Jannach et al. \cite{jannach2015recommenders} mention in their research that recommender systems aim at boosting recommendations from the long tail of the item popularity distribution, supposedly increase the sales of novel items. Several works try to provide both accurate with novel~\cite{Castells2015, clark2008, Vargas:2011:RRN} or diversified item recommendations~\cite{clark2008, Cheng2017}, where diversified item recommendation lists try to capture even more potential aspects of users' interest. In terms of MF, related work~\cite{Cremonesi2010, Yin_VLDB2012} has claimed that by increasing the number of latent factors of the basic matrix factorisation model~\cite{Koren2009}, we can more accurately recommend novel items.

Given a ranking of recommended items a so-called re-ranker can be employed to increase the novelty or diversity of the final recommendation list, no matter which algorithm generated it initially. Particularly the trade-off between precision and novelty are based on the two criteria: (i) the items' relevance to the user's preferences (e.g. how similar is the item to previous user's choices) and/or 
(ii) the item's contribution in diversifying the item recommendation list (how different is an item from those items are already placed in the top positions of the recommendation list). Thus, re-ranking approaches apply a secondary or post-processing ranking step before delivering the final item recommendation list to the target user.

A well-known re-ranking approach, which was originally proposed for ranking documents in the information retrieval (IR) domain, is Maximal Marginal Relevance (MMR)~\cite{Carbonell1998}. MMR chooses a document to be placed in the top positions of the ranking list, if it has the maximum similarity with the user's query/profile and also the minimum similarity to the previously suggested documents. Another re-ranking approach is known as xQuAD (eXplicit Query Aspect Diversification)~\cite{Santos2010}. This is a probabilistic model that takes under consideration both (i) the document's relevance probability and (ii) the aspect's diversity probability, towards a user's query/profile. Recently, xQuAD was adapted accordingly to consider within the same user profile different tastes/aspects by performing Sub-Profile Aware Diversification, denoted as SPAD \cite{KayaB17}. Another probabilistic re-ranking model~\cite{Vargas2014} represented items and the categories they belong to, as a linear combination of their global and local appearance probability over item and user profiles, respectively.

\color{black}

A different research direction in matrix factorisation formulates the item recommendation problem not as a rating prediction problem, but as a ranking problem using pairs of positive items (in the train set) and negative items (not in the train set) as pairwise input. For example,  Bayesian Personalised Ranking (BPR)~\cite{rendle2009bpr} optimises a simple ranking loss such as AUC (the area under the ROC-curve) and uses matrix factorisation as the ranking function, that can be optimised directly
using a stochastic gradient algorithm. Similarly to BPR, Ning and Karypis~\cite{ning2011slim, ning2012sparse} proposed  a set of Sparse LInear Methods (SSLIM), which involve an optimisation process to learn a sparse aggregation coefficient matrix based on both user-item purchase matrices and item side information.

 In the same direction, Wasilewski and Hurley~\cite{Hurley:2013, Hurley:2016} have proposed a matrix factorisation framework to trade-off between the accuracy of item recommendations and the diversity of the items in the recommendation list. Please notice that in Wasilewski and Hurley~\cite{Hurley:2013, Hurley:2016}, there is some discussion about the appropriate sign of their regularisation term.  In particular, the sign in front of the gradient term should be chosen as negative when maximisation of the regularisation term is required and positive when minimisation is required. They argue for a maximisation of their objective formula, such that the distance between item factors is small when the diversity is large. Moreover, TagiCoFi~\cite{zhen2009tagicofi} adds an additional term to the classic MF formula to employ the user similarities defined based on tagging information to regularise the MF procedure in order to make the user-specific latent feature vectors as similar as possible if the corresponding users have similar tagging history. In particular, TagiCoFi~\cite{zhen2009tagicofi} requires the additional factor to be minimised, such that the distance between item factors is small when their social distance is small.
Please notice that TagiCoFi aims at increasing the accuracy of recommendations and not on novelty and/or explainability of recommendations. Furthermore, it tries to minimise the distance between users in the latent factor space, which can be stressed as a similarity of our work with TagiCoFi~\cite{zhen2009tagicofi}. While \cite{zhen2009tagicofi} minimises the distance between two users in the latent space, we minimise the distance between a user and a novel/explainable item within this space.

In the following, we argue further why there is very small overlap between the work of  Wasilewski and Hurley~\cite{Hurley:2013, Hurley:2016} and our work, by identifying four important differences. The first is that similar to the previous approaches, their MF model computes the pair-wise ranking loss of the objective function (not the element-wise square loss like our methodology). In other words, our MF model is element-wise and predicts the missing values of the user-item rating matrix, whereas their model tries to optimise items' pairwise ranking. The second difference is that we are exploring the trade-off between item recommendation accuracy and item novelty, whereas they explored the trade off between item recommendation accuracy and item diversity. This difference is discussed further in the discussion section. The third difference is the fact that items' diversity by nature is not personalised, whereas in our case item's novelty is personalised. That is, in our case, users based on their experiences can consider the same exact item as more or less novel (i.e., differently), whereas in both of their models the diversity between any two or more items in a recommendation list is always the same for all users. The last difference is that we add two additional terms to our objective formula to minimise the distance between a user with a novel or/and explainable item in the latent space, whereas they add only one additional term to the classic MF objective formula and try to maximise the distance between these two items instead of the distance between a user and an item. In addition, we have to note that they separately try to maximise only by considering the regularisation term that captures how much items differ to each other based on their content by minimising their basic pair-wise ranking loss function. In other words, the reason that they use different signs for adding the gradient of their regularisation term to their update rule is based on the fact that they try to maximise or minimise only this regularisation term by following the goal to minimise the pairwise loss function.
\color{black}

In contrast to the aforementioned works of Abdollahi and Nasraoui~\cite{Abdollahi:2017}, Cremonesi et al.~\cite{Cremonesi2010}, Wasilewski and Hurley~\cite{Hurley:2013, Hurley:2016}, and Yin et al.~\cite{Yin_VLDB2012}, our proposed method incorporates two additional constraint terms (one for novelty and one for explainability) into the basic matrix factorisation formula. This additional information is taken from two external resources namely the user-item explainability matrix and the user-item novelty matrix, which will both be defined in the next section. While both novelty and explainability are seen as a key feature of recommendation utility in real scenarios, to our knowledge, there is no work on relating to both of them simultaneously and measuring their trade-offs.

\section{Item Explainability and Novelty}
\label{sec:measuring}

Based on the framework of Vargas and Castells\cite{ Castells2015, Vargas:2011:RRN}, who already defined the novelty of an item for a specific target user, we will use it to define the explainability of an item for a given user in this section. This enables us to measure besides accuracy also the degree of novel and explainable items an algorithm recommends to users. 

\color{black}

Table~\ref{Table:Symbols} summarises the symbols used in the following sections.

\scriptsize
\begin{table}[htb]
\centering
\begin{tabular}{cl} \toprule
{\bf Symbol}       & {\bf Definition} \\ \midrule
$k$                & number of nearest neighbours \\
$L_u$				& recommendation list for user $u$\\
$N$                & size of recommendation list \\
$NN$($u$)          & nearest neighbours of user $u$ \\
$P_{\tau}$         & threshold for positive ratings \\
$I$          & domain of all items \\
$U$          & domain of all users \\
$R$			& domain of the rating scale \\
$u,v$              & some users \\
$i,j$              & some items \\
${I}_u$       & set of items rated by user $u$ \\
${U}_i$       & set of users rated item $i$ \\
$r_{u,i}$          & the rating of user $u$ on item $i$ \\
$\overline{r}_u$   & mean rating value for user $u$ \\
$\overline{r}_i$   & mean rating value for item $i$ \\
$p_{u,i}$          & predicted rate for user $u$ on item $i$ \\
$\left|T\right|$   & size of the test set \\
$n$                & number of training users \\
$m$                & number of items \\ 
$N_i$				& novelty of item $i$ \\
$E_{ui}$			& Explainability of item $i$ for user $u$ \\
\bottomrule
\end{tabular}
\caption{Symbols and definitions.}\label{Table:Symbols}

\end{table}
\normalsize

\subsection{Personalised Item Explainability}

There are different ways for explaining a recommendation~\cite{HKR00}. Herlocker et al.~\cite{HKR00}
explored 21 different interfaces for explaining collaborative
filtering (CF) recommendations. They demonstrated that the ``User" style is persuasive in supporting explanations. To prove this, they conducted a survey with 210 users of the MovieLens recommender system, demonstrating that explanations can improve the acceptance of CF systems.

\begin{figure}[tbh]
\centering
\fbox{\includegraphics[scale=0.35]{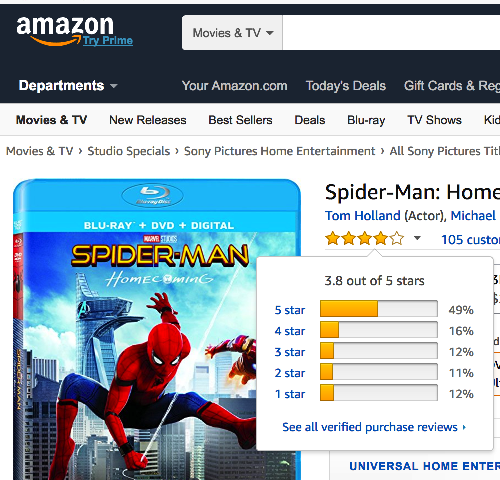}}
\caption{User style explanation example of the online retailer Amazon.}
\label{amazon_user}
\end{figure}

In this paper, we consider the ``user" style of explanations~\cite{BM05}, which provides explanations based on what items similar others to a target users have favourably rated. In other words, the  ``user" style of explanation is based on other users performing similar actions (buying items, rating items, etc). Please notice that the qualitative user study of Herlocker et al.~\cite{HKR00} has shown already the merit of the ``user" style of explanation, and many other user studies followed, showing its effectiveness~\cite{BM05, papadimitriou2012generalized}. 
Thus, the burden of proof that this type of explanation is insightful for users is out of scope of this paper. The merit of this explanation style is also indicated by the fact that several big product retailers, such as Amazon, Zalando, etc., adopted this user style of justifications: ``Customers who bought item $X$ also bought items $Y, Z, \ldots$".  For example, as shown at the right-bottom of Figure~\ref{amazon_user}, Amazon's interface displays a histogram of
users' ratings for an item.  

\begin{figure}[!htb]
\begin{framed}
We recommend you Movie 1 because your neighbours' ratings for this movie are the following:
\vspace{2em}

\centering
	\begin{tabular}{lc}

  \hline

    Rating 	& Number of Neighbours\\
    \hline
    \includegraphics[width = 1em]{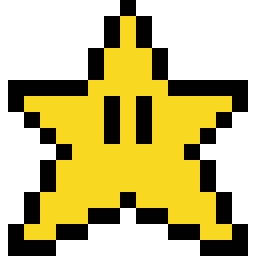} & 0 \\
    \includegraphics[width = 1em]{Star} \includegraphics[width = 1em]{Star}
    & 0 \\
    \includegraphics[width = 1em]{Star} \includegraphics[width = 1em]{Star}\includegraphics[width = 1em]{Star}
    & 0 \\
    \includegraphics[width = 1em]{Star} \includegraphics[width = 1em]{Star}\includegraphics[width = 1em]{Star}\includegraphics[width = 1em]{Star}
    & 10\\
    \includegraphics[width = 1em]{Star}\includegraphics[width = 1em]{Star}\includegraphics[width = 1em]{Star}\includegraphics[width = 1em]{Star}\includegraphics[width = 1em]{Star} & 23\\
    \hline
	\end{tabular}
    
    \end{framed}
    
    \caption{An explanation interface that also uses the explainability power of nearest neighbours for a target user.}
    \label{user_stars1}
\end{figure}

\begin{figure}

\begin{framed}

\centering
We recommend you Movie 2 because your neighbours' ratings for this movie are the following:

\begin{tikzpicture}
    \begin{axis}[
        width  = 8cm,
        height = 4cm,
        major x tick style = transparent,
        axis y line* =left,
    	axis x line* =bottom,
        ybar=2*\pgflinewidth,
        bar width=10pt,
        ymajorgrids = true,
        ylabel = {Number of Neighbours},
         xlabel = {Ratings},
        symbolic x coords={1-star, 2-stars, 3-stars, 4-stars, 5-stars},
        xtick = data,
        nodes near coords,
    nodes near coords align={vertical},
    %every node near coord/.append style={rotate=90, anchor=west},
        scaled y ticks = false,
        enlarge x limits=0.2,
        ymin=0,       
        legend cell align=left,
        legend columns=2,
        legend style={
                anchor=north east,
                column sep=0.5ex,
                scale = 3
        }
         legend image code/.code={%
                    \draw[#1, draw=none] (0cm,-0.1cm) rectangle (0.6cm,0.1cm);
                }
    ]
    
         \addplot[style={bblue,fill=bblue,mark=none}]
             coordinates {
             (1-star,1)
             (2-stars,2) 
             (3-stars,7) 
             (4-stars,14)
             (5-stars,9)};

        %\legend{0-20\%,20-40\%,40-60\%,60-80\%, 80-100\%}
    \end{axis}
\end{tikzpicture}

\end{framed}

\caption{A second explanation interface that uses the explainability power of nearest neighbours for a target user.}
\label{user_stars2}
\end{figure}
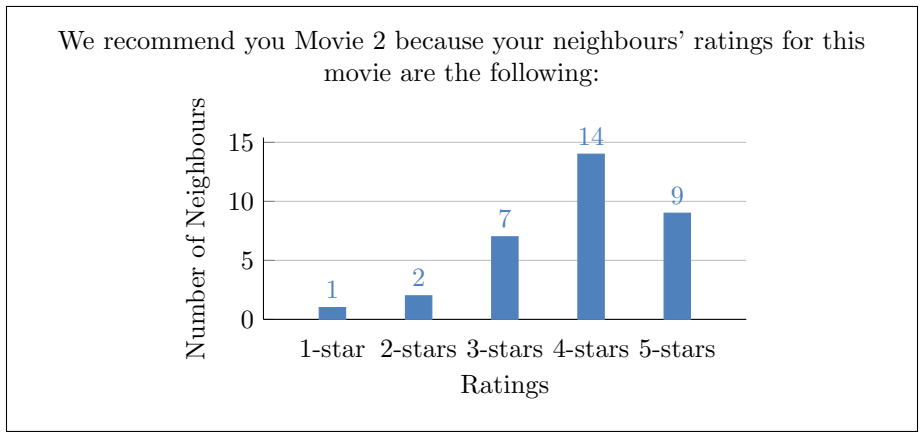

We can also see two variations of the same ``User'' style of explanation interface in Figures~\ref{user_stars1} and Figures~\ref{user_stars2}. In Figure~\ref{user_stars1} the user is
presented with the exact ratings his neighbours have entered while
in Figure~\ref{user_stars2} the interface displays a histogram of
neighbours' ratings for the recommended item.

%%%%%%%%%%histogram explanation

\subsubsection{Defining how explainable is a Recommended Item }

When computing the explainability power of an item $i$ for a user $u$ based on her/his neighbourhood preferences, we first have to identify the similar neighbours of a target user $u$ from the original user-item rating matrix. In particular, we can use the Pearson correlation (Equation~\ref{Eq:Pearson}), which measures the
similarity between two users, $u$ and $v$. %, where $r_{x,i}=R(x,i)$.

\footnotesize
\begin{equation}
\label{Eq:Pearson}
{\rm sim}(u,v) = \frac{\displaystyle\sum_{\forall i \in I_u \cap I_v}
(r_{u,i}-\overline{r}_u)(r_{v,i}-\overline{r}_v)}
{\sqrt{\displaystyle\sum_{\forall i \in S}
(r_{u,i}-\overline{r}_u)^2}\sqrt{\displaystyle\sum_{\forall i \in S}
(r_{v,i}-\overline{r}_v)^2}}, 
\end{equation}
\normalsize

where $r_{u,i}$, $r_{v,i}$ are the ratings on item $i$ of user $u$ and user $v$, respectively. $I_u$ and $I_v$ are the set of items rated by user $u$ and $v$, respectively. Means $\overline{r}_u$, $\overline{r}_v$ are the mean ratings of $u$ and $v$ over their co-rated items. Equation~\ref{Eq:Pearson} takes into account only the set of items, $S$, that are {\it co-rated} by both users.

Then, we set a number $k$ of nearest neighbours for the target user $u$, and inside this neighbourhood, we define $NN^k(u)_{i,r}$ as the number of $k$ nearest neighbours of target user $u$ who have given rating $r$ on item $i$. Please notice that $r$  $\in$ [1..R] rating scale.

Then, for a user $u$, who is recommended an item $i$, we compute how explainable item $i$ is for $u$ by measuring in the identified neighbourhood (i.e. the $k$ most similar user profiles) how frequently item $i$ has been highly rated. In other words, we construct a user-item explainability matrix $E$ that holds the explainability power of an item $i$ for a user $u$ as given in Equation~\ref{eq:explainability1}:

 \begin{equation}
	E_{u,i} = \displaystyle\sum_{\substack{\forall r \in R \\ r \geq P_{\tau}}} r * |NN^k(u)_{i,r}|,
	\label{eq:explainability1}
\end{equation}
 
where $NN^k(u)$ is the set of $k$ nearest neighbours of a user, and $R$ being the set of all different ratings in the rating scale. Please also note that  $NN^k(u)_{i,r}$ is the set of nearest neighbours of target user $u$ who gave a ``positive'' rating $r$ (above $P_{\tau}$ threshold) on item $i$ in the past. As expected, the values of matrix $E$ should be higher for items that are highly rated from many users and low in the opposite case. 

Please notice that using Equation 2, we are able to define the explainability power of an item for a target user based on the weighted frequency sum of ratings for a particular item among the nearest neighbours of a given user.

\color{black}

Based on Equation~\ref{eq:explainability1} we can measure how explainable an item is for both examples given in figures~\ref{user_stars1} and ~\ref{user_stars2}, respectively. 
As far as Figure~\ref{user_stars1} is concerned, the explainability power for justifying the recommendation of \text{Movie 1} is 159 ($4 \times 10 + 5 \times 23 = 159$), while for Figure~\ref{user_stars2} the explainability power for justifying the recommendation of Movie 2 is 127 ($1 \times 1 + 2 \times 2 + 3 \times 7 + 4 \times 14 + 5 \times 9 = 127$). Thus, following this approach Movie 1 is supposedly more explainable than Movie 2 since neighbours were more likely to give higher ratings while still the overall number of ratings is the same for this example. In the next section we will present results from an initial user study that confirms these assumptions underlying our proposed explainability measure.

\subsubsection{Evaluating the explainability of a recommendation list}
\label{def:explainability}

 As already discussed in the user-based style of explanation, we want to understand if the recommended items are also explainable  w.r.t. the explainability distribution of items. 

% Again comparing the algorithm's performance only with the visual representation  can be complicated. 
 
 For instance, Abdollahi and Nasraoui~\cite{Abdollahi:2017} measure the explainability of an item recommendation list $L_u$, which is provided to user $u$, as the ratio of the number of explainable items inside $L_u$ to the size of the number $N$ of recommended items (i.e., Mean Explainability Precision), as shown in Equation~\ref{mep}:

\begin{equation}
	MEP = \frac{1}{|U|} \times \sum_{u \in U}\frac{|\{i:i\in L_u, E_{u,i} > 0\}|}{N},
	\label{mep}
\end{equation}

where $U$ is the set of all users. As will be shown experimentally later, the main drawback of \textit{MEP} is the fact that it can not distinguish a more explainable item from a less explainable one. Moreover, it is obvious that for small values of the explainability threshold, $P_T$, MEP is almost always equal to 1, even if $L_u$ consist of items with small explainability power. Thus, to obtain a finer level of granularity in the aforementioned problems, we adopt the notion of Explainable normalized Discounted Cumulative Gain ($E\textnormal{-}nDCG_u$), which also takes under consideration the relative position of the recommended items inside $L_u$. 

The first step in the computation of the not normalised $E\textnormal{-}DCG_u$ is the creation of the \textit{gain vector}. In our case, the gain vector for each item $l$ in $L_u$, consists of its  explainability power ($E_{u,l}$), as it is defined in Equation~\ref{eq:explainability1} for the user style of explanation.

The second step in the computation of $E\textnormal{-}DCG_u$ applies the \textit{Discounted Cumulative Gain} to the aforementioned gain vector, as shown in Equation~\ref{DCG}. 

\begin{equation}
E\textnormal{-}DCG_u = E_{u,l_1} + \sum_{i=2}^N \frac{E_{u,l_i}}{log_2i}
\label{DCG}
\end{equation}

Based on Equation~\ref{DCG}, we discount the gain at each item rank inside $L_u$ to penalise items, which are recommended lower in the ranking, reflecting the additional user effort in order to reach the lower ranks and appreciate the corresponding explanation\cite{clark2008}. 

The third step is to normalise the $E\textnormal{-}DCG_u$ against the ``ideal'' gain vector. In our case, the ``ideal'' gain vector considers all recommended items in $L_u$ as having maximum explainability power, $E_{max}$. That is, all recommended items in $L_u$ are considered as being rated with the maximum rating $max$ (e.g., with 5 in the rating scale 1 to 5), from all neighbours of the target user and item, for the two explanation styles, respectively. Thus, the ideal E-IDCG is calculated as:
\begin{equation}
	E\textnormal{-}IDCG = E_{max} + \sum_{i=2}^{N} \frac{E_{max}}{log_2i},
\end{equation}
\begin{gather}
E_{max} = max \times |NN|,
\end{gather}
where $|NN|$ is the size of the neighbourhood that is used for explaining a recommendation. Finally,  $E\textnormal{-}nDCG_u$ is the ratio between E-DCG to E-IDCG, as shown in Equation~\ref{eq:finalnDCG}.

\begin{equation}
E\textnormal{-}nDCG_{u} = \frac{E\textnormal{-}DCG_u}{E\textnormal{-}IDCG}
\label{eq:finalnDCG}
\end{equation}

%The $E\textnormal{-}nDCG$ of the system is the average value of all the $E\textnormal{-}nDCG_u$ for every $u\in U$.

\subsection{Personalised Item Novelty}
\label{sec:novelty}

The premise of recommender systems is to suggest users non-popular items that match their interests, i.e. to make {\it novel} item recommendations. By doing this, businesses can increase their returns, since these novel items usually might have higher margins, or lower customer churn rates, since users would get bored and disappointed by receiving trivial recommendations of already known popular items. In the following, we will therefore define the \textit{novelty} of a recommended item and how to measure the novelty of a recommendation list. %strong claim, reference required. 

\subsubsection{Defining how Novel is a Recommended Item}
\label{def:novelty2}

Related work~\cite{Castells2015} in recommender systems has proposed several definitions of item novelty. The popularity-based novelty definition~\cite{Castells2015}, for instance, also known as global long-tail novelty, focuses on discovering relatively unknown items coming from the long-tail of the item popularity distribution. However, by using the popularity-based novelty, we can capture only the novelty of an item over the whole crowd of the recommender system, but we miss to address the novelty for a particular person (i.e., personalised item novelty).

Differently to the case of popularity-based item novelty, we  will use a distance-based model~\cite{Castells2015, Vargas:2011:RRN}, also known as unexpectedness, where item novelty is defined by a distance function between the target item $i$ from the set of items $I$ and the set of items $I_u$ that a user has already interacted with (the user's past experience). We can formulate this novelty as shown in Equation~\ref{eq:novelty3}:

 \begin{equation}
	N_{u,i} = \frac{\displaystyle\sum_{\substack{\forall j \in I_u}}  d(i,j)}{|I_u|}
	\label{eq:novelty3}
\end{equation}

Please note, that the distance between two items can be also considered as the complement (i.e. $d(i,j)=1-sim(i,j)$) of any similarity measure (cosine-based, Jaccard coefficient, etc.) in terms of the item features (i.e., the category that an item belongs to, the features of an item, etc.) or the user's item categories profile~\footnote{To capture the interaction between users and the item categories they have interacted with, we can build a user-category profile, composed  of the user-item rating profile and the item-category profile (e.g., their dot product).} (i.e. the item categories that a user presumably prefers). For example, in news recommendation, we know the category to which an article belongs to (i.e., politics, sports, etc.). Thus, when we recommend an article about sports to a user, who has already seen a lot of stories about sports, this will obviously not be considered to be as novel as for a person, who has never seen an article about sports before. In order to capture how novel a topic category is for a user, we can use Equation~\ref{eq:topic_coverage}, which is based on the well-known subtopic recall metric (S-recall)~\cite{Castells2015}, but adjusted to our case scenario:

 \begin{equation}
	N_{u,c} = \frac{1}{|\{i \in I_u : i \text{ belongs to category c}\}|}
    \label{eq:topic_coverage}
\end{equation}

where $i$ is an item and $C$ is the set of all topic categories. Thus, when a user $u$ has interacted with many items that belong to the same category $c$, then any additional item from this category will be considered to be less novel for user $u$.

\subsubsection{Evaluating the Novelty of a Recommendation List}
\label{sec:noveltylist}

In this Section, we will define an integrated way of measuring  the novelty of a recommendation list, which can be used either for popularity-based or personalised item novelty. For a user $u$ who is recommended $N$ items, we adopt the following definition of novelty of a  recommendation list of items $L_u$~\cite{ Castells2015, Vargas:2011:RRN}:

\begin{equation}
N_{L_u}=   \frac{1}{N}\displaystyle\sum_{\forall i \in L_u}  N_{u,i}
\label{Eq:novelty3} 
\end{equation}

where $N_{u,i}$ is the novelty of item $i$ for the target user $u$ that can be any of the \textit{item novelty} models shown in column 3 of Table~\ref{tab:noveltymetrics}. Please note that the last column of  Table~\ref{tab:noveltymetrics} shows the context of the user's experience. For the popularity-based model it consists of all the users that interacted with item $i$ (i.e., $U_i$), whereas for the distance-based model, it consists of the items that only user $u$ has interacted with in the past (i.e. $I_u$).

\begin{table*}

\caption{Different item/category novelty definitions and the metric they result into.}
\centering
\label{tab:noveltymetrics}
\begin{tabular}{llccc}
\toprule
\multirow{2}{*}{Metric} 				& 	\multirow{2}{*}{Ref.}				&		Item Novelty  		& Ideal Novelty  &	\multirow{2}{*}{Context}\\ 
&&($N_{u,i}$)& ($N_{max}$)\\
\midrule

Popularity-based	& 	\multirow{2}{*}{\cite{Castells2015,Vargas:2011:RRN}}	&		\multirow{2}{*}{$ -log\frac{|U_i|}{|U|}$} &		\multirow{2}{*}{$ -log\frac{1}{|U|}$} 	& \multirow{2}{*}{$U_i$} \\ 
(Global Long-Tail)	& 		&			&  &\\ 

Distance-based		&	\multirow{2}{*}{\cite{Ge:2010:BAE:1864708.1864761}}					&		\multirow{2}{*}{$\frac{\sum_{\substack{\forall j \in I_u}}  d(i,j)}{|I_u|}$}&		\multirow{2}{*}{$d_{max}$}	& \multirow{2}{*}{$I_u$} \\
(Unexpectedness)	&					&								&	&\\ 

Topic-Coverage		&	\multirow{2}{*}  {\cite{Castells2015}}					
      &		\multirow{2}{*} 
{$\frac{1}{|\{i \in I_u : i \in c^\dagger\}|}$}& 	\multirow{2}{*}{$1$} & \multirow{2}{*}{$C_u$} \\
(S-Recall)	&					&								&	&\\ \bottomrule
\end{tabular}
\begin{minipage}[t]{0.9\textwidth}
$^\dagger$ c is a Category.
\end{minipage}
\end{table*}
\normalsize
A drawback of Equation~\ref{Eq:novelty3}, which captures the novelty of a recommendation list $L_u$, is that it cannot penalise the fact that a less novel item is ranked at a better position in $L_u$ than another more novel item. In analogy to the E-nDCG defined in Section~\ref{def:explainability}, we therefore introduce the following Novel - normalized Discounted Cumulative Gain ($N\textnormal{-}DCG_u$).

$N\textnormal{-}DCG_u$ considers the relative position of the recommended novel items inside $L_u$. We compute the $N\textnormal{-}DCG_u$ gain vector for each item $l$ in $L_u$, which consists of its novelty ($N_{u,l}$), as it is defined in column three of Table~\ref{tab:noveltymetrics}, and apply the \textit{Discounted Cumulative Gain} to the aforementioned gain vector, as shown in Equation~\ref{eq:N-nDCG1}. 

\begin{equation}
N\textnormal{-}DCG_u = N_{u,l_1} + \sum_{i=2}^N \frac{N_{u,l_i}}{log_2i}
\label{eq:N-nDCG1}
\end{equation}

In Equation~\ref{eq:N-nDCG1}, we discount the gain at each item's rank inside $L_u$ to penalise items, which are recommended lower inside $L_u$. Next, we normalise the $N\textnormal{-}DCG_u$ against the ``ideal'' gain vector. In our case, the ``ideal'' gain vector considers each item in $L_u$ as having maximum novelty, $N_{max}$. $N_{max}$ is different for each item novelty model, as shown in column four of  Table~\ref{tab:noveltymetrics}. Please notice that the $d_{max}$ variable, which is shown in the fourth column of Table~\ref{tab:noveltymetrics} depends on the range of the used distance function. Thus, the ideal N-IDCG is calculated as:
\begin{equation}
	N\textnormal{-}IDCG = N_{max} + \sum_{i=2}^{N} \frac{N_{max}}{log_2i},
\end{equation}

Finally, the $N\textnormal{-}nDCG_u$ is the ratio between $N\textnormal{-}DCG_u$ to N-IDCG:

\begin{equation}
N\textnormal{-}nDCG_{u} = \frac{N\textnormal{-}DCG_u}{N\textnormal{-}IDCG}
\end{equation}

\section{User Study}
\label{sec:userstudy}

Collaborative explanations that justify recommendations by showing rating summaries such as the total number of similar users and the mean value of their ratings helps users to understand the reasoning behind recommendations and thus ease the decision making. 

Next we will therefore present results from an initial user study that researched if and how collaborative explanations building on rating summary statistics actually guide users' choices.

Between January and February 2018 a group of 73 people was invited to participate in a controlled experiment to explore how users perceive different configurations of user style explanations. We base our analysis on the Choice-Based Conjoint (CBC)~ \cite{louviere2010discrete} methodology. In CBC designs, products (a.k.a., {\it profiles}) are modelled by sets of categorical or quantitative \textit{attributes}, which can have different \textit{levels}. In CBC experiments, participants have to repeatedly select one profile from different \textit{sets of choices}, which nicely matches real-world recommendation settings where users are confronted with lists of items alongside with explanations. The participants were presented with the following hypothetical situation, and had to repeatedly select one profile from choice sets containing 3 items each. The choice task was accompanied with the following briefing of participants:

\textit{``Assume that you find yourself in the situation that you need to make a choice between three movies to watch on a movie platform. 
These three movies are equally preferable to you with respect to all other movie information you have access to (title, plot, actors etc.).
Other similar users' ratings are aggregated and summarised by their number of ratings, the mean rating value and their distribution.
Therefore, we would like to know your choice, by solely considering these rating summary statistics.''}

\subsection{Results on Users' Favourite Attributes}

Detailed results for estimating users' preference weights are presented in Table~\ref{tab:resultsmultilogit}. The fourth column of Table~\ref{tab:resultsmultilogit} shows the coefficients (or preference weights) for each level of the two attributes, where the base levels (i.e. small number of ratings and low mean ratings) have been constrained to be zero. The fifth and the sixth column of Table~\ref{tab:resultsmultilogit} report the standard errors and the P-values for the respective levels of each attribute.

%%%%%%%%%%%%%%%%%%%%%%%%%%%%
%%%%%%%%%%%%%%%%%%%%%%%%%%%%
%%%%%%%%%%%%%%%%%%%%%%%%%%%%
%%%%%%%%%%%%%%%%%%%%%%%%%%%%
%%%%%%%%%%%%%%%%%%%%%%%%%%%%
\begin{table}[t]
\centering
\caption{Estimates for the multinomial logit model.}

\begin{tabular}{lllrl}
\toprule
Attribute & Level & Level Value & \multicolumn{2}{l}{Coefficient} \\
\midrule
\multirow{3}{*}{\shortstack[l]{Number\\ of Ratings}} & Large & L3: \textit{2970} &0.53 (0.14)	& *** \\
								& Medium & L2: \textit{560} & 0.13 (0.16) \\
                            	& Small & L1: \textit{290} & -\\

\midrule 

\multirow{3}{*}{Mean Rating} & High & L3: \textit{3.6} & 2.77 (0.27) &	***	 \\
								& Average & L2: \textit{3.3} & 1.09 (0.29)	&	*** \\
                            	& Low & L1: \textit{2.9} & - \\         

\bottomrule
\end{tabular}
\label{tab:resultsmultilogit}

\begin{minipage}[t]{0.8\textwidth}
\footnotesize
 \textit{Note:} $^{***}	p<0.001$;	 $^{**} p<0.01$;	 $^{*} p<0.05$. Dashes (-) are the baseline levels. The estimated coefficients are the change in log odds of choosing a particular level rather than the baseline. The values in parentheses are estimated standard errors.
\end{minipage}
\end{table}

\begin{figure}[h]
\centering
\includegraphics[scale=.5]{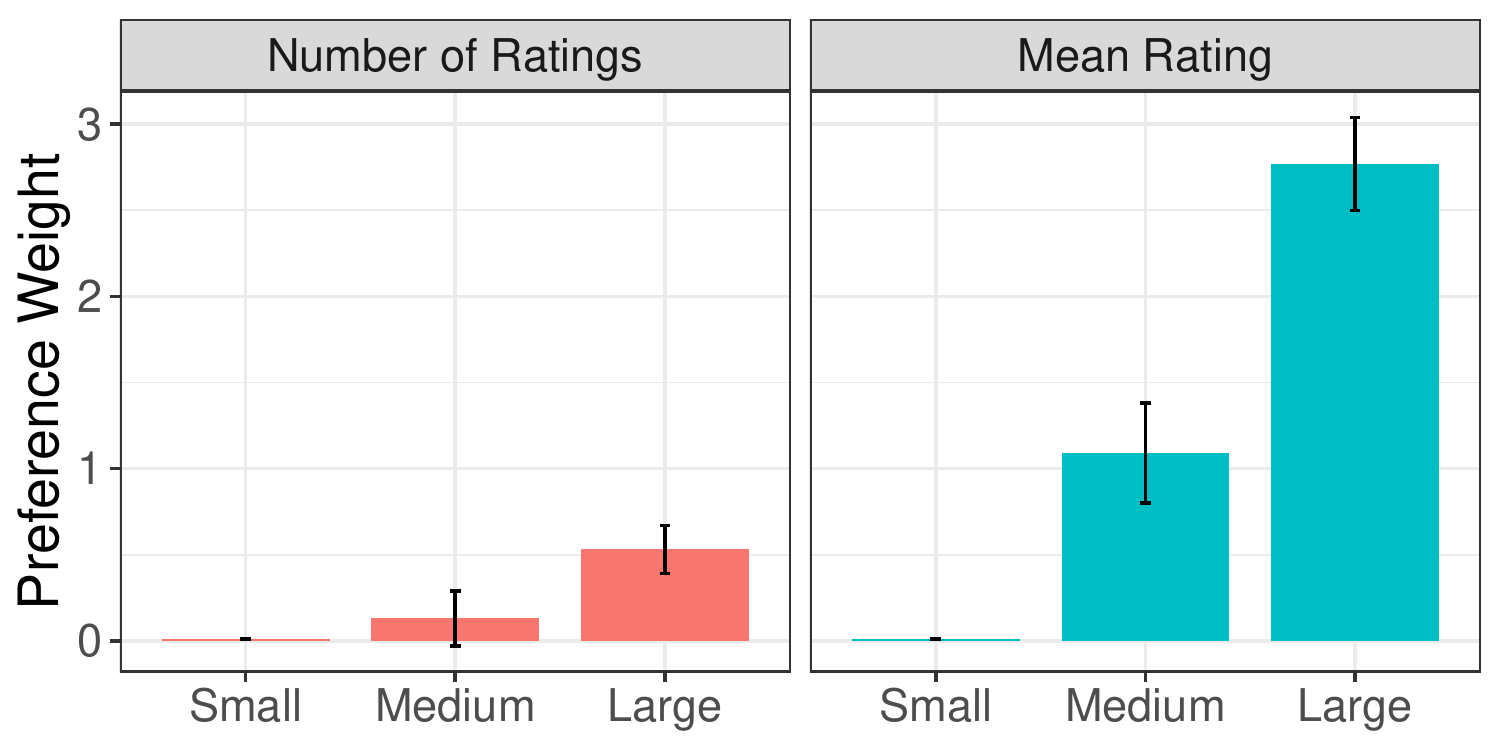}
\caption{Preference weights of the multinomial logit model.}\label{fig:multiLogit}
\end{figure}

Figure~\ref{fig:multiLogit} visually depicts the preference weights of the multinomial logit model for each level of the two selected attributes (i.e. total number and mean of ratings). As expected, there was a general {\it higher is better} tendency for the two attributes - i.e., users prefer bigger numbers of ratings and higher mean values. 
There was a clear and statistically significant preference relation over the three levels for mean rating values. However, in terms of the total number of ratings, users did not seem to care that much. The large number of ratings was statistically significant, and clearly preferred over the other two levels, but between the medium and small level, the P-value was above the threshold of .05 (cmp. Table~\ref{tab:resultsmultilogit}) and thus no statistically noticeable difference in the users' perception existed. 

Thus, the conjoint analysis clearly supports our modelling assumptions, that higher mean rating values have a strong influence on the explainability of a rating summary statistic.
Please notice that in future, we want to comprehensively exploit these findings in our algorithm to provide more effective explanations. 

\subsection{Results about Users' Favourite Style}

Next, we measured users' perception over different explanation types based on the origin of the ratings used (i.e., if the ratings derived from friends or from similar users). In other words, we have used two styles of explanations, based on the origin of the ratings. That is, if the target user takes an explanation based on the ratings of other users that are similar or based on the ratings of her/his friends in social networks like Facebook or Instagram. The first approach can be considered the ``User' style of explanations (denoted as style $A$), and the second one the ``Friend' style (denoted as style $B$).

We assumed that explanation style $B$ will
be the users' favourite on due to homophily. However, our results as illustrated in
Table~\ref{Tab:survey2} have a surprising result: the relative importance in terms of percentage of style $B$ is only $perc_A$ = 20.5\%, whereas $perc_B$ is 79.5\%. This contradicts our initial assumption that explanation style $B$ would be users'
favourite choice, see Table~\ref{Tab:survey2}.
% We run paired t-tests with the null hypothesis $H_0(perc_{diff} = 0)$.
%These confidence intervals are estimated with a 0.05 significance level, and reject our assumption that explanation style $B$ is the users' favorite choice.

\footnotesize
\begin{table}[!htb]

\centering
\begin{tabular}{ ccc ccc }
\toprule

Freq$_A$ & Perc$_A$ & Conf.Int$_A$ & Freq$_B$ & Perc$_B$ & Conf.Int$_B$ \\
 \midrule 
 58 &  79.5\% & $(67.8\%,91.1\%)$ & 15 & 20.5\% & $(8.9\%,32.2\%)$ \\
 \bottomrule
\end{tabular}

\caption{Results of the user survey for the user and the friend style of explanations.}
\label{Tab:survey2}
\end{table}

\normalsize

\begin{figure}[!htb]
\begin{center}
\includegraphics[scale=0.5]{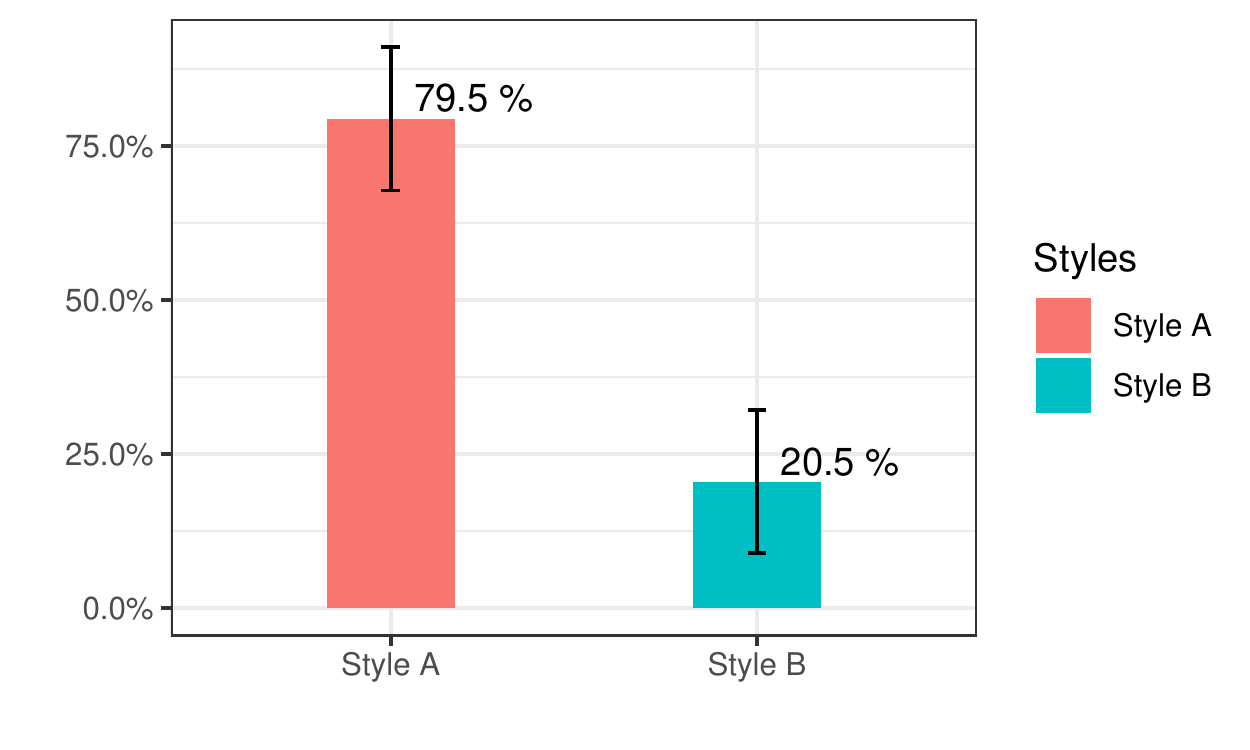}
\end{center}
 \caption{Relevant preference and standard deviation of  explanation styles $A$ and $B$.}
 \label{Fig:frienddiagram}
\end{figure}

Finally, Figure~\ref{Fig:frienddiagram} visually represents the preference for the 
explanation styles $A$ and $B$ including confidence intervals. 

Thus, the outcome of the initial user study clearly supports our algorithmic focus on collaborative explanations exploiting ratings from users' nearest neighbours.

\section{Proposed NEMF method}\label{objfun}

As mentioned before we address novelty in analogy to explainability and propose an algorithmic framework to trade-off between explainability and novelty in matrix factorisation. In this Section, we present our method and the steps that took us there.

\subsection{Explainable Matrix Factorization}
\label{sec:emf_manh}

As far as the items' explainability is concerned, to provide more explainable items, Abdollahi and Nasraoui~\cite{Abdollahi:2017} proposed Explainable Matrix Factorisation (EMF), where they put an additional constraint on the classic regularised matrix factorisation formula as shown in Equation~\ref{eq:adding_constraint}:

\begin{center}
\begin{gather}
 G_{explainable} =  \sum_{i, j \in R}(r_{ij} - {u_{i}v^T_{j}})^2 + 
\frac{\beta}{2} {(\|u_{i}\|^2 + \|v_{j}\|^2)} + 
\nonumber \\
\lambda \|u_{i}-v_{j}\|^2  E_{ij},
\label{eq:adding_constraint}
\end{gather}
\end{center}

where $R$ is the set of ratings for user $i$ on item $j$, $\|u_{j}\|^2,  \|v_{j}\|^2$ are the $L_2$  regularisation terms and $E_{ij}$ holds the information of how explainable is item $j$ for user $i$. Please notice that  $\|u_{i}-v_{j}\|$ constraints the representations of the user/item vectors in the latent space, in such a way so that they are close to each other (i.e., their difference is close to zero), in order to minimise the objective function.

Then, to minimise the objective function $G$, they compute the error of the difference among the real and the predicted rating values of items by using a numerical method, such as \textit{Gradient Descent}, and by applying the following update rules:

\begin{gather}
u'_{i} =  u_{i} + \eta(2 (r_{ij} - {u_{i}v^T_{j}}) v_{j} - \beta u_{i} - \lambda (u_{i} - v_{j }) E_{ij})
\nonumber \\
v'_{j} = v_{j} + \eta(2 (r_{ij} - {u_{i}v^T_{j}}) u_{i} - \beta v_{j} + \lambda (u_{i} - v_{j}) E_{ij})
\end{gather}

where $\eta$ is the
learning rate, whose value determines the step rate for detecting the minimum. Parameter $\beta$ is used to control the magnitudes of the user-latent feature and item-latent feature vectors, whereas parameter $\lambda$ controls the explainability matrix.

In this Section, we reformulate the objective function of EMF of Equation~\ref{eq:adding_constraint}, to overcome the problem of Euclidean distance's metric over high dimensional spaces. That is, Euclidean distance (L2 norm) places more emphasis on outliers, computing a much higher weight (i.e., by  squaring the difference) for them, which may dominate other smaller weights computed for other normal data points. In contrast, Manhattan distance (L1 norm) tries to reduce all errors equally since its gradient has constant magnitude. We have experimentally identified that Manhattan distance (L1 norm) is preferable to Euclidean distance (L2 norm) for the case of high dimensional data. Based on the aforementioned argument, we change the additional soft explainability constraint of Equation~\ref{eq:adding_constraint}, by using the Manhattan distance, as shown by Equation~\ref{equation:manhattan}:

\begin{center}
\begin{gather}
 G_{explainable} =  \sum_{i, j \in R}(r_{ij} - {u_{i}v^T_{j}})^2 + 
\frac{\beta}{2} {(\|u_{i}\|^2 + \|v_{j}\|^2)} + 
\nonumber \lambda \|u_{i}-v_{j}\|  E_{ij},
\label{equation:manhattan}
\end{gather}
\end{center}

Then, to minimise the objective function $G_{explainable}$, we apply the following update rules:

\begin{gather}
u'_{i} =  u_{i} + \eta \cdot (2 \cdot (r_{ij} - {u_{i} \cdot v^T_{j}}) \cdot v_{j} - \beta \cdot u_{i} - \lambda \cdot sgn(u_i - v_j) \cdot E_{ij})
\nonumber \\
v'_{j} = v_{j} + \eta \cdot (2 \cdot (r_{ij} - {u_{i} \cdot v^T_{j}}) \cdot u_{i} - \beta \cdot v_{j} - \lambda \cdot sgn(u_i - v_j) \cdot E_{ij}),
\label{eq:16}
\end{gather}

where $u_i - v_j$ $\neq$ 0 and  the explainablity regularisation term $\lambda$ $>$ 0. Please notice that $sgn()$ is the signum or else the sign function and for a vector $x$ = [$x_1$,$\ldots$, $x_d$], $sgn(x)$ = [$sgn(x_1)$,$\ldots$, $sgn(x_d)$], and $sgn(x_j)$ = 1 when $x_j$ $>$ 0, and $sgn(x_j)$ = -1 when $x_j$ $<$0, and $sgn(x_j)$ = 0 when $x_j$ = 0.

\color{black}

\subsection{Novel Matrix Factorisation}
\label{sec:nmf}

For the personalised novelty, in analogy to the EMF-manhattan case, which we described in the previous section, we add an additional soft constraint for novelty into the classic regularised matrix factorisation formula as shown in Equation~\ref{eq:novelMF}:

\begin{center}
\begin{gather}
 G_{Novel} =  \sum_{i, j \in R}(r_{ij} - {u_{i}v^T_{j}})^2 + 
\frac{\beta}{2} {(\|u_{j}\|^2 + \|v_{j}\|^2)} + 
\nonumber \delta \|u_{i}-v_{j}\|  N_{ij},
\label{eq:novelMF}
\end{gather}
\end{center}

where $\delta$ controls the novelty vector and $N_{ij}$ holds the information of how novel is item $j$ and the update rules of are in accordance to those that are shown of the EMF-manhattan. Then, to minimise the objective function $G_{novel}$, we use the \textit{Gradient Descent} method.

\subsection{Novel and Explainable MF}

Consequently, we want to define the objective function such that we can push for more novel and more explainable items into top-$N$ recommendation lists at the same time with minimal losses in the recommendation accuracy.  To do this, we incorporated the additional constraint terms from Sections~\ref{sec:emf_manh} and ~\ref{sec:nmf} into the objective function as shown in the following:

\begin{center}
\begin{gather}
\label{eq:nemf}
 G_{expl \& novel} =  \sum_{i, j \in R}(r_{ij} - {u_{i}v^T_{j}})^2 + 
\frac{\beta}{2} {(\|u_{i}\|^2 + \|v_{j}\|^2)} + 
\nonumber \\
 ||u_{i}-v_{j}|| (\lambda E_{ij} + \delta N_{ij}),
\end{gather}
\end{center}

The new update rules change as follows:

\begin{gather}
u'_{i} =  u_{i} + \eta(2 (r_{ij} - {u_{i}v^T_{j}}) v_{j} - \beta u_{i} - sgn(u_i - v_j)(\lambda E_{ij} + \delta N_{ij}))
\nonumber \\
v'_{j} = v_{j} + \eta(2 (r_{ij} - {u_{i}v^T_{j}}) u_{i} - \beta v_{j} - sgn(u_i - v_j)(\lambda E_{ij} + \delta N_{ij})),
\label{18}
\end{gather}

where $u_i - v_j$ $\neq$ 0, the explainability regularisation term $\lambda$ $>$ 0, and the novelty regularisation term $\delta$ $>$ 0. An attractive property of L1 norm is that it induces sparsity into the predicted model and thus, it can create more compact and interpretable models. In particular, to induce more sparsity into our predicted model, we need to increase the value of parameter $\lambda$. That is, with L1 norm our model can attain as many zero elements as possible~\cite{morup2007multiplicative} giving the simplest and most interpretable and explainable solution to account for the data.
In other words, L1 norm has the tendency to select sparse solutions (i.e., few nonzero components) for the predicted model, and it is particularly effective for high-dimensional data with many non-correlated user and/or non co-rated item features. Thus, L1 norm identifies those item/user features with zero coefficients and effectively discards them. In summary, the main advantage of L1 norm over L2 norm is not always the one of performance in terms of accuracy, but the fact that it is highly interpretable and explainable~\cite{morup2007multiplicative}. Henceforth, we call this method Novel and Explainable Matrix Factorisation (NEMF). Please notice that MF, EMF, and NMF methods, are just simplified special cases of NEMF and can be easily derived from it. 

\color{black}

%\begin{equation}
%    sign(x) = \left\{
%  \begin{array}{@{}ll@{}}
%    1, & \text{if}\ x > 0, \\
%    -1, & \text{if}\ x < 0,\\
%    0, & \text{if}\ x = 0
%  \end{array}\right.
%\end{equation}

%-----------------------------------------------------------------------%

\section{Experimental Results}
%\subsection{Evaluation Metrics}
\label{sec:results}

In this Section, we experimentally compare our approach NEMF with other recommendation algorithms. We use the Explainable Matrix Factorisation (EMF)\cite{Abdollahi:2017} algorithm, and the Matrix Factorisation~\cite{Koren2009} algorithm (MF) for the comparison. Moreover, we will use the Maximal Marginal Relevance re-ranker (MMR)~\cite{Carbonell1998} combined with the MF algorithm, such that we have in our experiments also a variation of MF, which focuses on providing novel item recommendations. In particular, for re-ranking with MMR the item recommendation list provided from classic MF algorithm, we adapt the following greedy objective function of Sa{\'u}l Vargas \cite{vargas2011new}, as shown by Equation \ref{eq:mmr}:

\begin{gather}
argmax [(1-\lambda) * \hat{r}_{norm}(u,i) + \lambda * avg_{j \in L_u} (1-sim(i,j))],
\label{eq:mmr}
\end{gather}
where $\hat{r}_{norm}(u,i)$ is the normalised predicted rating of user $u$ over item $i$. It is normalised in [0,1] scale, such that it can be combined with the Jaccard similarity\footnote{The Jaccard similarity is particularly adequate for binary data. In this case we are considering the similarity in the context of the topic coverage.} (see the second term of Equation \ref{eq:mmr}), which measures the dissimilarity of items. In particular, it measures the average similarity of an item with all other items which have already taken a position inside the $L_u$ recommendation list, which is to be re-constructed for user $u$. As can be shown by the $\lambda$ parameter of Equation \ref{eq:mmr}, there is a trade-off between how relevant an item is being considered by a user, and how much this item differs from the items, which have been already included inside the currently constructed recommendation list. In our experiment we kept the trade-off $\lambda = 0.5$. 
We also distribute the code for reproducing the experiments online.\footnote{\url{https://github.com/proton35/explainable_MF}}

\color{black}

\subsection{Data Sets}

Our experiments are performed on two datasets, MovieLens 100K (ML100K) and MovieLens 1ML (ML1M)~\cite{Harper2015}. ML100K consists of 100,000 ratings assigned by 943 users on 1,682 movies. 
ML1M contains 1,000,209 anonymous ratings of approximately 3,900 movies made by 6,040 users. The range of ratings is between 1(bad)-5(excellent).  
The datasets also contain the movie genres, whereas a movie may belong to one or more genres, as shown in Table~\ref{tab:datasets}. Please note that since the data sets contain information about the genres of the movies, we have used the distance-based item novelty for our experiments, thus the novelty of an item may not be the same for different users. 
\begin{table}

\caption{Datasets.}
\centering
\begin{tabular}{lcc}
\toprule
Characteristic 	& ML-100K	&	ML-1M	\\ \midrule
\# of ratings	& 100,000 &	1,000,209 \\ 
\# of users & 943 & 6,040 \\ 
\# of items & 1,682 & 3,952\\ 
\# of genres & 19 & 18 \\ 
Average \# of genres per item & 1.7 & 1.6\\ 
Rating's domain & [1,5] & [1,5]\\ \bottomrule
\end{tabular}
\label{tab:datasets}
\end{table}

\subsection{Experimental Protocol and Evaluation}

Our evaluation considers the division of items of each target
user into two sets: (i) the training set $E$$^T$ is treated
as known information and, (ii) the probe set $E$$^P$ is used
for testing and no information in the probe set is allowed to be
used for prediction. It is obvious that, $E = E$${^T}
\cup E$${^P} $ and $E$${^T} \cap E$${^P} = \oslash$. Therefore, for a target user we generate the recommendations
based only on the items in $E$$^T$.

Except for the metrics E-nDCG and N-nDCG, that are introduced in Sections~\ref{def:explainability} and~\ref{sec:noveltylist}, respectively, we use precision, and nDCG metrics as classic accuracy performance measures for item recommendations.

Finally, in our experiments we will also present MEP, which is presented in Section~\ref{def:explainability} for comparisons reasons, as proposed by the work of Abdollahi and Nasraoui~\cite{Abdollahi:2017}.

We perform all experiments with $4$-fold cross validation,  with a training-test split percentage, 75\%-25\%.  The default size of the recommendation list $N$ is set to 10, except for cases that denoted differently.  The presented measurements, based on two-tailed t-test, are statistically significant at the 0.05 level. All algorithms predict the items of the target users' in the probe set.

We evaluate NMF, EMF, and NEMF by varying the novelty and explainability regularisation terms, while we keep the other parameters fixed to values denoted next. For ML100K, the number of latent factors and the learning rate $\eta$ for all algorithms is set to 80 and 0.001, respectively, whereas the number of users used for explaining a recommendation is set to 10. For ML1ML, the number of latent factors for all algorithms is set to 50, parameter $\eta$ is set to 0.001 and the number of users used for explaining a recommendation is set to 10.

\color{black}

\subsection{Sensitivity analysis of NMF}

\begin{figure}[!htb]
    \centering
    \begin{subfigure}[!htb]{1\textwidth}
        \centering
        \includegraphics[scale=.5]{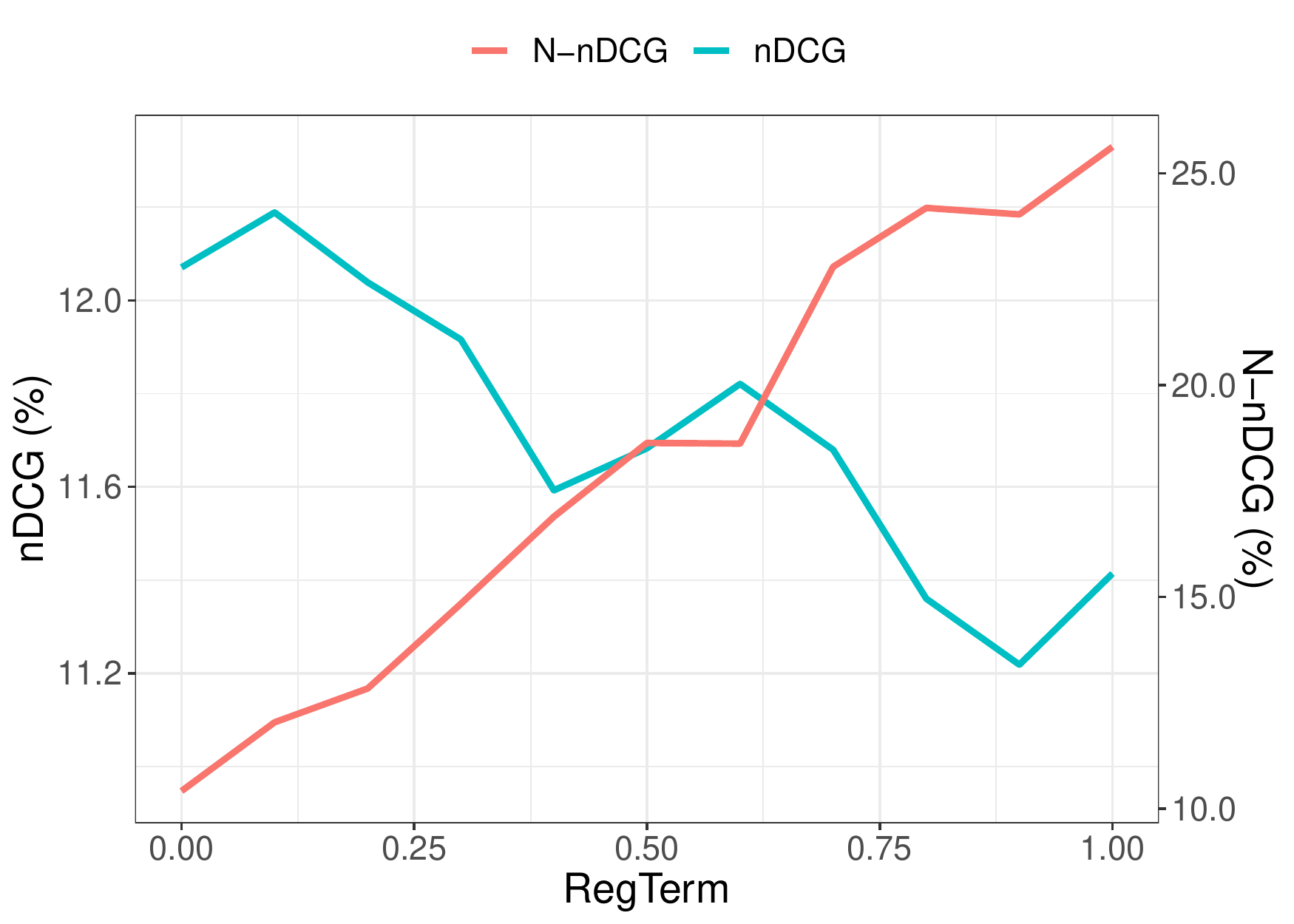}
        \caption{}
        \label{sensitivity_NMF_ML100K}
    \end{subfigure}%

    \begin{subfigure}[!htb]{1\textwidth}
        \centering
        \includegraphics[scale=.5]{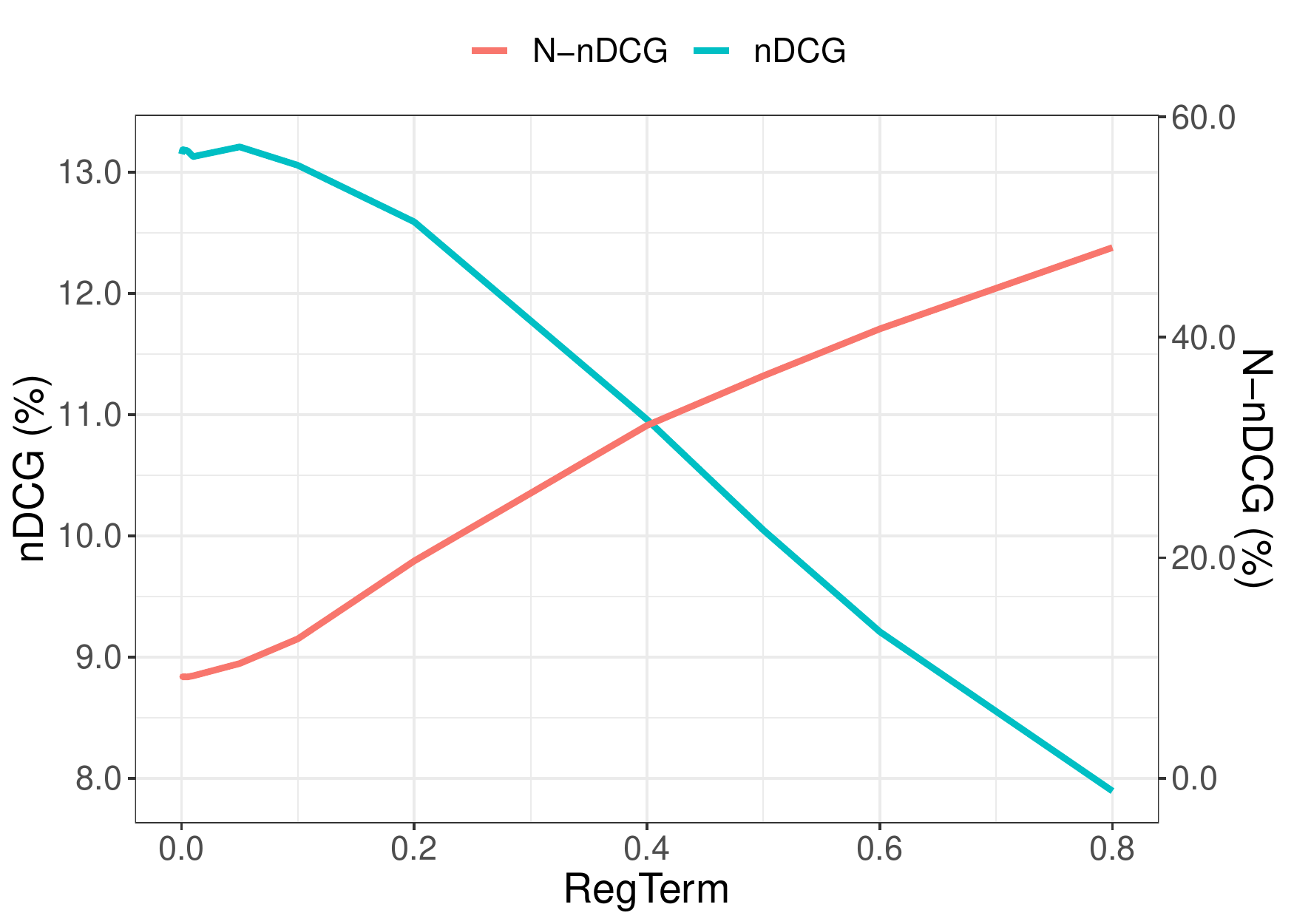}
        \caption{}
        \label{sensitivity_NMF_ML1M}
    \end{subfigure}
    \caption{Sensitivity of NMF to changes of the novelty regularization term for (a) the ML100K and (b) the ML1M data sets.}
\end{figure}

In this Section, we want to explore how the performance of NMF is affected when we increase the impact of the regularisation term $\delta$, which controls novelty in Equation~\ref{eq:novelMF}. Both Figures~\ref{sensitivity_NMF_ML100K} and~\ref{sensitivity_NMF_ML1M} show that as we increase $\delta$, N-nDCG increases, whereas nDCG decreases. That is, as we increase $\delta$, NMF  recommends more novel items but the recommendation accuracy drops drastically, which indicates that novelty and precision accuracy are negatively correlated. 

On ML100K, we achieved the best novelty, in terms of N-nDCG, when parameter $\delta$ equal to 1. As can be seen from Figure~\ref{sensitivity_NMF_ML100K}, novelty (N-nDCG) and accuracy (nDCG) balance at $\delta = 0.4$, where the two lines cross.
On ML1M, the best novelty was achieved for $\delta$ equal to 0.8. As before, we can see in  Figure~\ref{sensitivity_NMF_ML1M}, novelty and accuracy balance at $\delta = 0.4$, where the two lines cross.

\color{black}
This trade-off is something that we will try to balance experimentally later with our proposed NEMF method. 

\subsection{Sensitivity Analysis of EMF}

In this Section, we want to explore how the performance of EMF in terms of providing explainable and accurate recommendations is affected, as we increase the impact of the regularisation term $\lambda$, which controls explainability in Equation~\ref{equation:manhattan}.

\begin{figure}[!htb]
    \centering
    \begin{subfigure}[!htb]{1\textwidth}
        \centering
         \includegraphics[scale=.5]{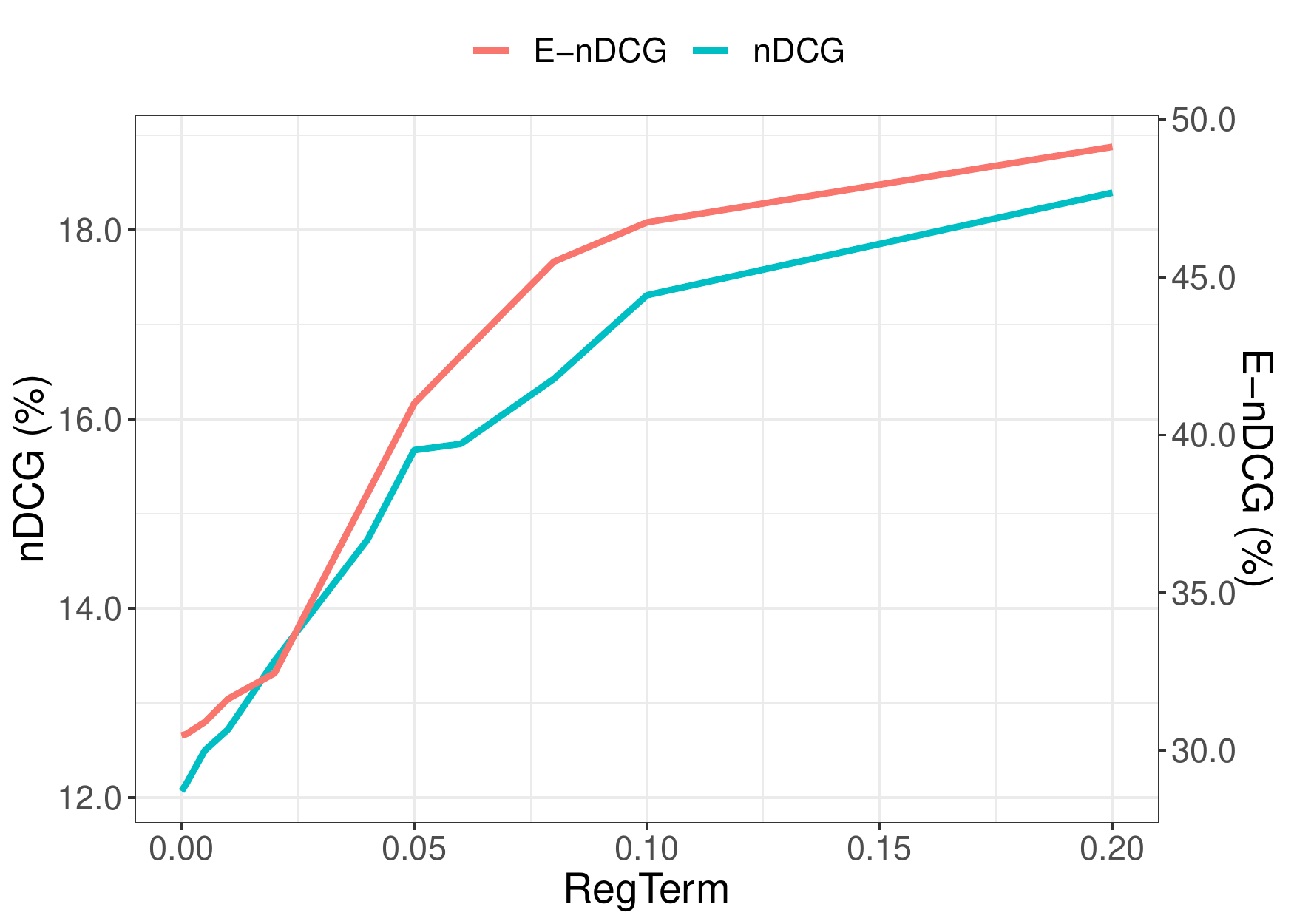}
        \caption{}
   \label{fig:sensitivity_explain_ML100K}
    \end{subfigure}%

    \begin{subfigure}[!htb]{1\textwidth}
        \centering
         \includegraphics[scale=.5]{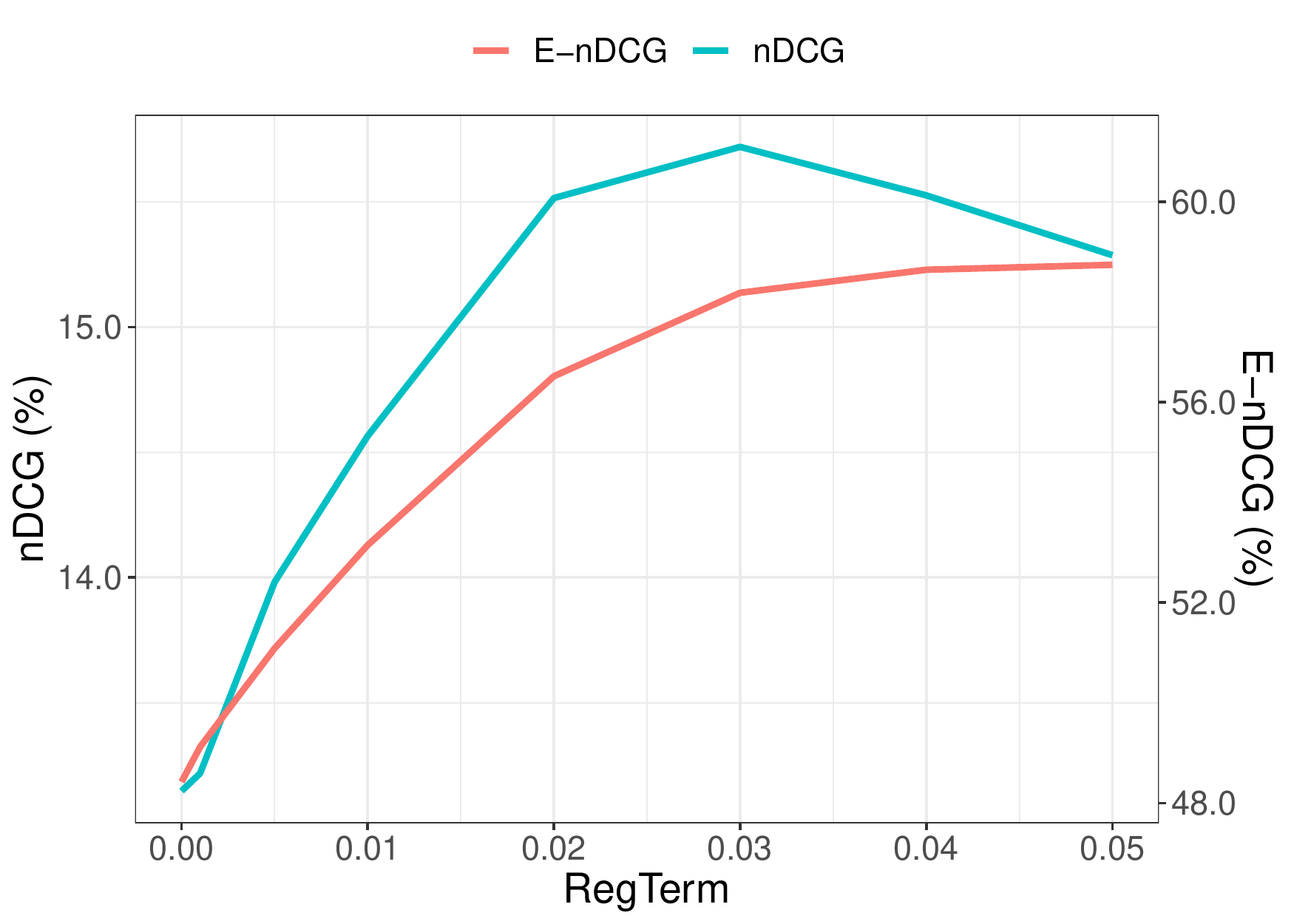}
        \caption{}
    \label{fig:sensitivity_explain_ML1M}
    \end{subfigure}
    \caption{Sensitivity of EMF to changes of the explainability regularisation term for the (a) ML100K and (b) ML1M datasets.}
\end{figure}

Both Figures~\ref{fig:sensitivity_explain_ML100K}
and~\ref{fig:sensitivity_explain_ML1M} show that, up to a point, as we increase $\lambda$, both E-nDCG and nDCG increase too. That is, as we increase $\lambda$, EMF  recommends more explainable items and the recommendation accuracy continues to increase, which means that explainability and precision accuracy are positively correlated. 

On ML100K we achieved best precision and explainability for $\lambda = 0.2$, while for ML1M the best explainability was achieved with $\lambda = 0.03$.

\color{black}

%explainable MF

Thus, by recommending explainable items, you might recommend the more popular ones, which is presumably a major reason for getting increased accuracy performance. However, these types of recommendations of popular items, even if accurate, might be perceived as boring by users due to their lack of serendipity and novelty. This type of trade-off we want to balance with our proposed NEMF method.

 \subsection{Sensitivity Analysis of NEMF}

\begin{figure}[!htb]
    \centering
    \begin{subfigure}[!htb]{1\textwidth}
        \centering
       %\fbox{ \includegraphics[scale=0.55]{3d-nemf_1}}
        \includegraphics[scale=0.4]{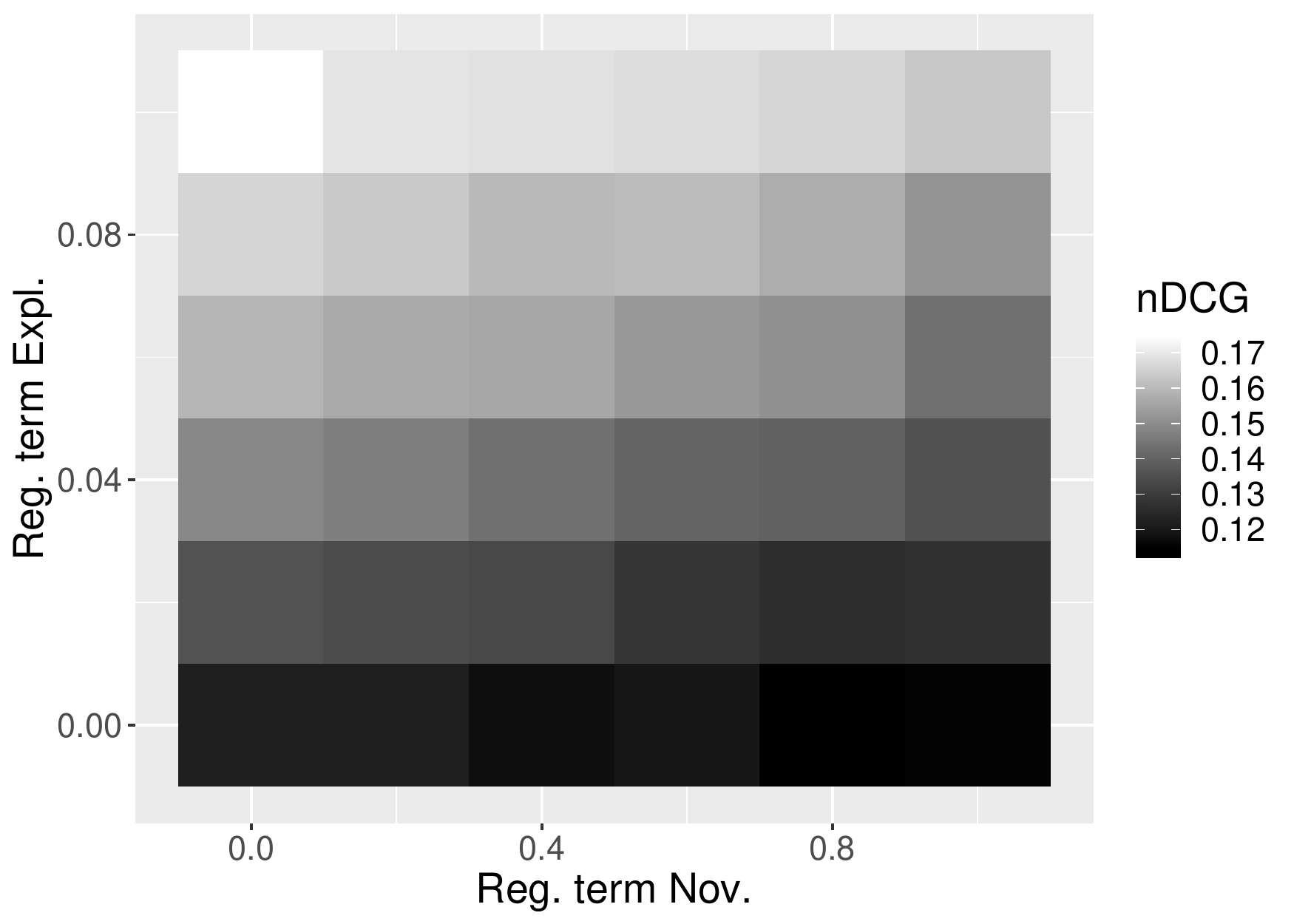}
        \caption{}
   \label{fig:sensitivity_nemf_ML100K}
    \end{subfigure}%

    \begin{subfigure}[!htb]{1\textwidth}
        \centering
       % \input{plots/nemf_1ml.tex}
         %\fbox{\includegraphics[scale=0.55]{3d-nemf_2}}
        \includegraphics[scale=0.4]{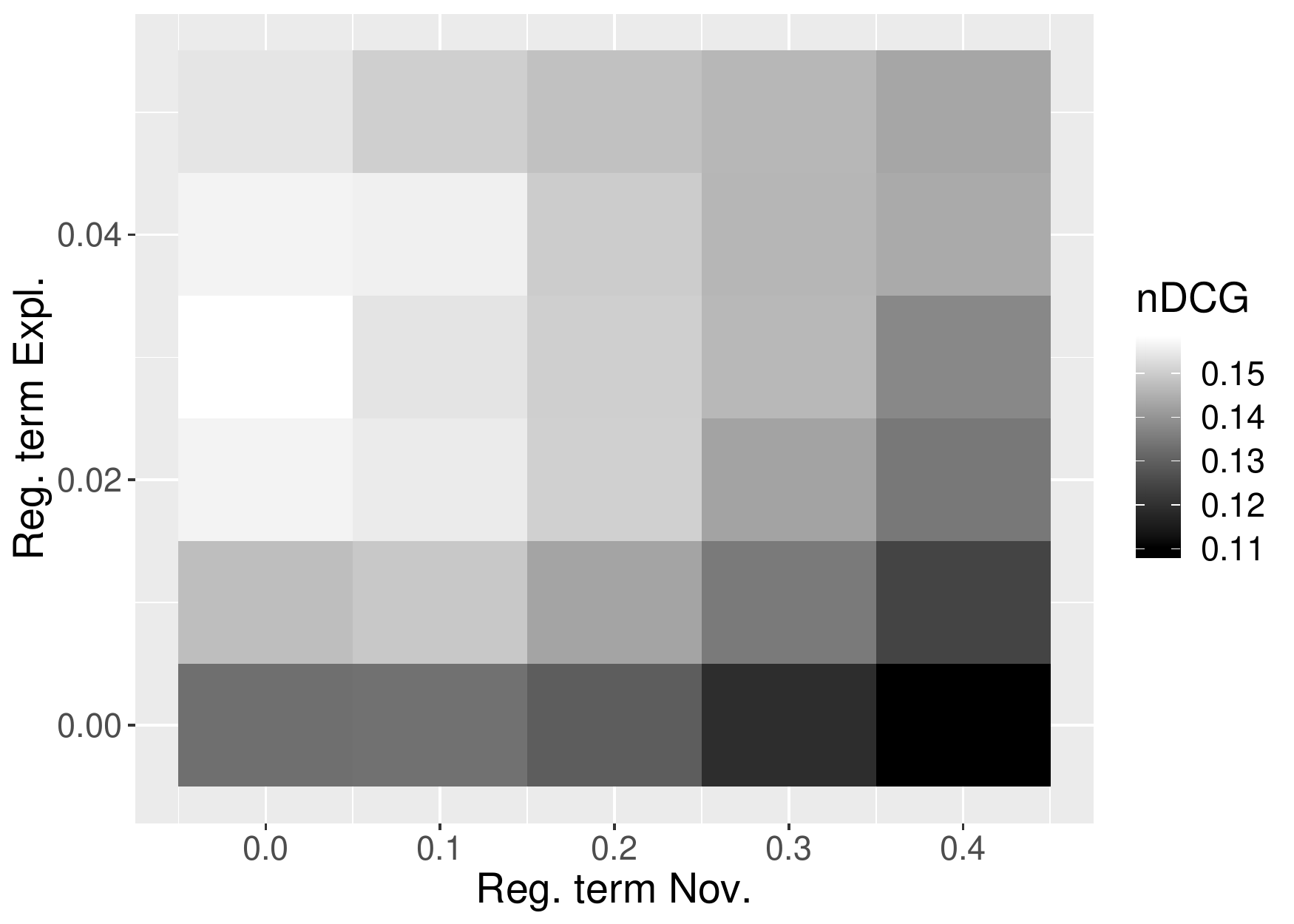}
        \caption{}
    \label{fig:sensitivity_nemf_ML1M}
    \end{subfigure}
    \caption{Sensitivity of the NEMF in terms of nDCG to changes of novelty and explainability regularisation terms on the (a) ML100K and (b) ML1M datasets.}
\end{figure}

In this Section, we want to explore how NEMF performs in terms of providing novel, explainable and accurate recommendations, when we increase the impact of the regularisation terms $\delta$ and $\lambda$, controlling novelty and explainability, respectively, in Equation 18. 
In terms of accuracy, as shown in Figure~\ref{fig:sensitivity_nemf_ML100K} and Figure~\ref{fig:sensitivity_nemf_ML1M}, respectively for ML100K and ML1M, we change simultaneously the weights of both regularisation terms of explainability and novelty.

Thus, we can observe how the recommendation accuracy in terms of nDCG varies as the other two parameters change. The brightness of the cells denotes the nDCG, brighter colours denote higher scores of the nDCG, while darker colours are for lower scores of nDCG. The two axes denote the corresponding weight values of the explainability and novelty regularisation terms. 

\color{black}
In terms of novelty and explainability performance of NEMF, we also run experiments to identify their best possible values. However, we have to note that it is impossible to determine the parameters' combination that achieves for all three metrics (i.e. accuracy, novelty and explainability) the maximum values. Thus we need to compromise in order to achieve reasonable results for all three target metrics.

  \subsection{Comparison with other methods}

In this Section, we compare our NEMF against MF, NMF, and EMF-Euclidean and EMF-Manhattan in terms of their accuracy (with precision and nDCG), explainability (with E-nDCG and MEP), and novelty (with N-nDCG) performance. 

Tables~\ref{tab:comparison_ML100K} and~\ref{tab:comparison_ML1M} give the performance results for all algorithms on the ML100K and the ML1M datasets, respectively, when we provide top-10 item recommendations. For both data sets, the best accuracy and explainability performance is attained by EMF-Manhattan, which is a method that is proposed in this paper. However, EMF-Manhattan does not perform well in terms of novelty. As expected, the best novelty is attained by NMF, but with severe losses in terms of precision/nDCG. As shown, only NEMF (see in the last row of both Tables~\ref{tab:comparison_ML100K} and~\ref{tab:comparison_ML1M}) is able to recommend high accuracy in combination with both good levels of novel and explainable item recommendations.

\begin{table}[!htb]
\centering
\caption{Algorithms' Recommendation Performance at top-10 recommended items on ML100K.}
\begin{tabular}{lccccc}
\toprule
\textbf{Algorithm}	&	Prec.	&	nDCG	&	MEP	&	E-nDCG	&	N-nDCG\\ 
\midrule
MF	&	11.02 \%	&	12.07 \%	&	80.26 \%	&	30.47 \%	&	10.40 \% \vspace{0.5em}\\ 
MF + \textit{MMR} & 10.88 \% & 12.23 \% & 78.23 \% & 29.52 \% & 13.07 \% 
\\  %tradeoff=0.3
NMF & 10.52 \% & 11.41 \% & 89.45 \% & 33.68 \% & \textbf{25.62} \%
 \vspace{0.5em}\\ 
%MF + MMR(expl.) & \\ \hline
EMF-$L_2$ & 11.28 \% & 12.39 \% & 83.16 \% & 31.76 \% & 8.74 \%
% & 7.9\%	&	8.1\%	&	43.1\%	&	6.9\%	&	10.2\%
\\ 
EMF-$L_1$ &	\textbf{14.34} \%& \textbf{17.01} \% & \textbf{98.47} \%& \textbf{46.69} \% & 5.78 \%

\vspace{0.5em}\\

NEMF$^\dagger$ &  12.29 \% & 14.17 \% & 90.86 \% & 39.40 \% & 14.14 \%%expl 0.2 nov 1 

 \\ \bottomrule

\end{tabular}
\begin{minipage}[c]{0.9\textwidth}
\scriptsize
$^\dagger$ best solution to trade-off between novelty and explainability. In \textbf{Bold} the highest value per metric.
\end{minipage}

\label{tab:comparison_ML100K}
\end{table}
\begin{table}[!htb]
\centering
\caption{Algorithms' Recommendation Performance at top-10 recommended items on ML1M.}
\label{tab:user}

\begin{tabular}{lcccccc}
\toprule

\textbf{Algorithm}	&	Prec.	&	nDCG	&	MEP	&	E-nDCG	&	N-nDCG\\ \midrule

MF	& 11.80 \% & 13.15 \% & 93.64 \% & 48.42 \% & 9.15 \%
\vspace{0.5em} \\ 
MF + \textit{MMR} & 9.05 \% & 10.39 \% & 91.13 \% & 42.17 \% & 9.41 \%
\\ 

NMF & 7.96 \% & 7.89 \% & 95.94 \% & 41.60 \% & \textbf{48.14} \% 
\vspace{0.5em} \\ 

EMF-$L_2$	& 11.77 \% & 13.13 \% & 93.70 \% & 48.20 \% & 9.16 \%
\\ 
EMF-$L_1$	& \textbf{13.63} \% & \textbf{15.29} \% & \textbf{99.81}\% & \textbf{58.74} \% & 7.36 \%
\vspace{0.5em}\\ 

NEMF$^\dagger$ & 11.02 \% & 11.52 \% & \% 98.33  \% & 50.01 \% & 28.55 \%  
\\ \bottomrule

\end{tabular}
\label{tab:comparison_ML1M}
\begin{minipage}[c]{0.9\textwidth}
\scriptsize
$^\dagger$ best solution to trade-off between novelty and explainability. In \textbf{Bold} the highest value per metric.
\end{minipage}
\end{table}

\section{Discussion}

The idea of explaining recommendations based on the number of nearest neighbours may help users to better understand the relevance of items. Of course, when a user style of explanations is applied on a social network, it provides the number of friends of the target user as explanations. In such cases, the recommendation along with its explanation would be as follows: ``We recommend you Item 1 because it is highly rated by 14 of your friends''.  
However, in this paper, we have focused on the user style of explanations as also supported by our initial user study. Please note, that our proposed explainability and novelty evaluation metrics can be used to evaluate the performance of other explainable recommendation algorithms, too. However, based on the fact that in our paper the explainability/novelty of an item is defined based on how many neighbour users rated highly the recommended item, the creation of the explainability and novelty user-item matrices (i.e., Equations~\ref{eq:explainability1} and ~\ref{eq:novelty3}, respectively) would require adaptations in order to be applicable for other explanation styles/algorithms. For example, we need to change Equation~\ref{eq:explainability1} to apply our methodology to the item-based knn algorithm along with the so-called ``Item" style of explanation~\cite{BM05}, where the justifications are of the following form: ``Item $Y$ is recommended because you highly rated/bought item $X, Z, \ldots$". Thus, the system depicts those items i.e., $X, Z, \ldots$, that influenced the recommendation of item $Y$ the most.

To compute the explainability power of an item $i$ for a user $u$ based on her/his preferences on other items (high ratings/purchases in the past), we have to first find the similar items of the recommended item $i$, from the original user-item rating matrix. Then, we create the recommended item's neighbourhood, by getting the $k$ number of most similar items of $i$. That is, for an item $i$, which is recommended to a user $u$, we can define $NN^k(i)_{u,r}$ as the set of the $k$ similar items of $i$, which were also given from user $u$ a ``positive'' rating $r$ (above $P_{\tau}$)  in the past. Please notice that $r$  $\in$ [1..R] rating scale. Then, for a user $u$ that is recommended an item $i$, we can compute how well explainable is $i$ for $u$, by summing up the ratings that he has made on the $k$ most nearest (similar) items to $i$. Thus, we can construct a user-item explainability matrix $E$ that holds the explainability power of an item $i$ for a user $u$ as it is shown in Equation~\ref{eq:explainability}:

 \begin{equation}
	E_{u,i} = \displaystyle\sum_{\substack{\forall r \in R \\ r \geq P_{\tau}}} r * |NN^k(i)_{u,r}|,
	\label{eq:explainability}
\end{equation}
 
 and $NN^k(i)_{u,r}$ is the $k$ number of similar items to the recommended item $i$, which were also given from user $u$ a ``positive'' rating $r$ (above $P_{\tau}$ threshold)  in the past. Please note that we would need to make analogous adaptations in Equation~\ref{eq:novelty3}, such that we are able to capture how novel an item is for a user, in respect to the 'item' style of explanation.  As part of our future work, we want to explore NEMF also for the ``Item" style of explanations as well as other explanation styles/ algorithms.

\color{black}

\normalsize

In this paper, we focused on the novelty of a recommendation list, but we did not see the diversity of the recommendation list, where a diversified item recommendation list tries to capture more aspects of a user's interest. For measuring an item's novelty, we need at least one recommended item (i.e., we can just measure if an item is novel for a user), whereas for measuring the diversity of a recommendation list we need at least 2 recommended items (e.g., intra-list diversification with topic coverage). In particular, the intra-list diversification, which measures how diversified are the recommended items inside the list based on the percentage of coverage of different topics that are covered with the items of the recommendation list.  As a next step, we want to evaluate personalised NEMF in terms of the intra-list diversification on different data sets.

In a multi-objective optimisation scenario of matrix factorisation, finding acceptable trade-offs between aspects such as, for instance, accuracy, explainability or novelty is challenging and requires further research. While we do not provide the full answer w.r.t. how to trade-off these system objectives, the paper, however, sheds additional light on the relationships between these objectives and proposes an algorithmic mechanism to adjust the weighting of these different goals. Moreover, we have to mention that there are also other methods, in which explainability is not being exactly aligned to the prediction algorithm, such as the Local Interpretable Model-agnostic Explanations (LIME) \cite{Ribeiro0G16} approach, which does not depend on the type of data, nor on a particular type of comprehensible local predictor or explanation. The main intuition
of LIME is that the explanation may be derived locally from the records generated randomly in
the neighbourhood of the record to be explained, and weighted according to their proximity to it.

\color{black}

\section{Conclusions}
\label{sec:conclusions}
In this paper, we proposed a framework for matrix factorisation, denoted as NEMF, that simultaneously considers both the novelty and explainability of recommended items. Our empirical results have revealed the trade-off relationships between algorithmic accuracy, explainability and novelty. We have also experimentally shown that, MF, EMF and NMF approaches are simplified special cases of NEMF and can be easily derived from it. As future work, we plan to assess the users' perception for more explanation styles, e.g. item-based style of explanation. Moreover, we want to explore how to apply the proposed NEMF in the context of social networks, where the idea of explaining a recommendation using similar users will be replaced by the user's friends, which can support many more application scenarios on the Social Web.

\end{document}